\documentclass[10pt,journal]{IEEEtran}
\usepackage{url}
\usepackage{color}
\usepackage{subfigure}
\usepackage{graphicx}
\usepackage{epstopdf}
\usepackage{booktabs}
\usepackage[table,xcdraw]{xcolor}
\usepackage{array}
\title{Network as a Service: The New Vista of Opportunities}
\author{Junaid Qadir$^{1}$, Nadeem Ahmed$^{1}$, Faqir Zarrar Yousaf$^{2}$, Ali Taqweem$^{1}$\\
\normalsize $^{1}$School of Electrical Engineering and Computer Science, National University of Sciences and Technology (NUST), Pakistan\\
\normalsize $^{2}$NEC Laboratories Europe, Heidelberg, Germany\\
\normalsize Email: $\{$junaid.qadir, nadeem.ahmed, 10beeataqweem$\}$@seecs.edu.pk,
zarrar.yousaf@neclabs.eu}

\date{}
\begin{document}
\maketitle

\begin{abstract}

The networking industry, compared to the compute industry, has been slow in evolving from a closed ecosystem with limited abstractions to a more open ecosystem with well-defined sophisticated high level abstractions. This has resulted in an ossified Internet architecture that inhibits innovation and is unnecessarily complex. Fortunately, there has been an exciting flux of rapid developments in networking in recent times with prominent trends emerging that have brought us to the cusp of a major paradigm shift. In particular, the emergence of technologies such as cloud computing, software defined networking (SDN), and network virtualization 
are driving a new vision of `networking as a service' (NaaS) in which networks are managed flexibly and efficiently cloud computing style. These technologies promise to both facilitate architectural and technological innovation while also simplifying commissioning, orchestration, and composition of network services. In this article, we introduce our readers to these technologies. In the coming few years, the trends of cloud computing, SDN, and network virtualization will further strengthen each other's value proposition symbiotically and NaaS will increasingly become the dominant mode of commissioning new networks.
\end{abstract}


\section{Introduction}

With the proliferation of smart devices and high bandwidth applications, there has been a great upsurge in data traffic in recent times motivating the development of novel network architectures that allow greater flexibility while restricting the increased capital and operational expenditure (CAPEX/OPEX). The traditional network architecture---which is based mainly on interconnections of vertically integrated proprietary hardware with coupled control and data planes---offers little in terms of high-level sophisticated abstractions to program the network. This has limited operators in rolling out new services and in adapting to changing load demands and application requirements.

In order to address the shortcomings of the traditional networks, the emerging concept of `network as a service' (NaaS) aims at leveraging the concepts of cloud computing and virtualization to provide pluggable, scalable and application programmer interface (API) driven network management where the users can deploy and manage their networks as virtual logical networks decoupled from their physical instantiation on the underlying network. NaaS provides a networking framework that extends cloud computing's on-demand and self-service provisioning model to the network service provider affording users and operators the same benefits of cloud computing. 

The viability of NaaS is bolstered by the increasing role of modern warehouse-sized datacenters (DCs) as well as advancement in virtualization technology in terms of security, isolation and performance. In a DC, hundreds of thousands of powerful commodity servers are placed in racks connected by commodity switches. Virtualization technology enables the instantiation of multiple virtual machines (VMs) on a single server within a DC with the VMs being managed by the VM manager also known as a \emph{hypervisor}. The VMs provide a faithful imitation of the original server's interface, while ensuring inter-VM isolation, to the applications running on the VM. This enables a network operator to host network service/functions on VMs while taking advantage of the programmability features of VM cloning and mobility (allowing transport of VM snapshots from a busy server to any underutilized physical server). More importantly, NaaS enables the operator to virtualize the networking components in a manner analogous to the server virtualization and to provide a \emph{`virtual network'} (VN) abstraction that is akin to the VM abstraction. This decoupling is essential for VNs to afford the same operational benefits that we have come to expect from VMs. 

The aim of this article is to provide a tutorial overview of the various technologies that enable NaaS. According to the best of our knowledge, no such article exists in literature, and filling this void is the main contribution of this work. The remainder of this article is organized as follows: The notion of NaaS, and its main motivations, is described in section \ref{sec:whatisNaaS}. We discuss the deficiencies of traditional network virtualization techniques in section \ref{sec:traditional}, before introducing the enabling technologies of NaaS in section \ref{sec:EnablingTechnologies}. Finally, we conclude this article in section \ref{sec:conclusion}.

\section{What is Network as a Service (NaaS)?} 
\label{sec:whatisNaaS}

\begin{figure}[t]
\begin{center}
\includegraphics[width=.4\textwidth]{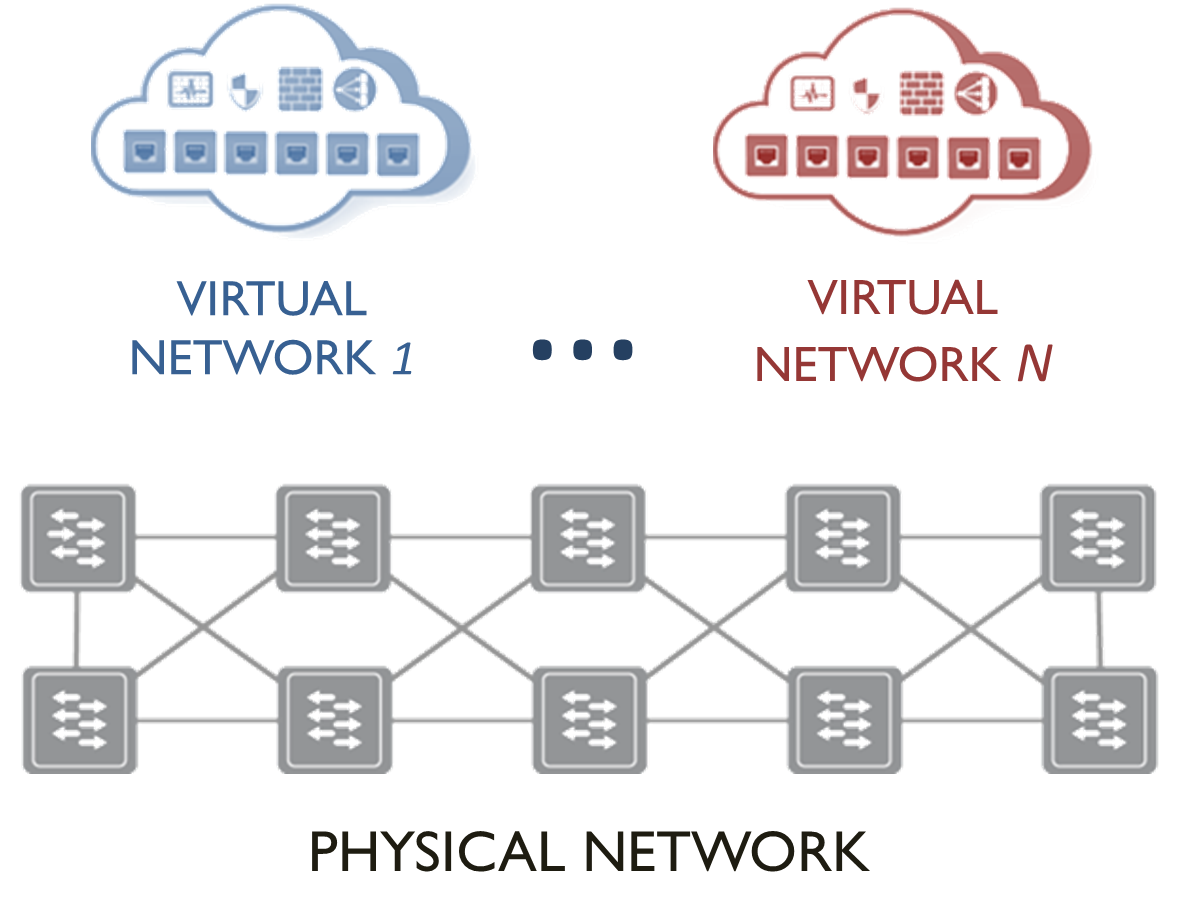}
\caption{In NaaS, virtual networks are logically decoupled from the underlying physical infrastructure substrate.}
\label{fig:vn}
\end{center}
\end{figure}

NaaS is in essence the `cloudification' of traditional networking. While VMs have unshackled applications from being tied to particular physical servers, traditional \emph{network virtualization} techniques---such as virtual LANs (VLANs), virtual private networks (VPNs)---do not offer an analogous VN abstraction that decouples the network from the physical infrastructure (refer to figure \ref{fig:vn}). NaaS is the vision of providing the VN abstraction as a service such that this VN abstraction can be instantiated, operated, cloned, moved, and repurposed as desired by the user in cloud computing style. 


\subsection{Virtual Network (VN) Abstraction}
\label{sec:VN}

The VN abstraction implies that the virtualized network be managed sans any manual interaction with physical assets by the network manager. Just like a VM is a software-container (encapsulating logical CPU, memory, storage, and networking) providing an interface identical to a physical machine to an application, a VN is also a software container encapsulating logical components (such as routers, switches, firewalls, etc.) which presents an interface identical to a physical network to network applications. Implementing this VN abstraction requires detachment of the virtual network from the physical infrastructure as well as isolation between multiple tenants sharing the same infrastructure. The VN abstraction allows great flexibility to IT managers as the physical network can now be managed as a \emph{`fabric'} offering some transport capacity that can be used, programmed, and repurposed as needed without requiring any change to the existing physical network or IP addresses, the networking workload, or the server virtualization technique. 

Since VNs have the same operational model as VMs, VNs should support dynamic creation, resizing, and mobility (within and between DCs). This is challenging in multi-tenant DCs where a tenant's VMs may reside in different servers, or even in different DCs \cite{koponen2014network}. To allow agile operations of cloud service providers (CSPs), it is important to have the ability of migrating VMs anywhere in the DC without restricting VM addressing to match the subnet boundaries of the underlying DC network. A key requirement for live migration is that a VM retains critical network state at its new location, including its IP and MAC address(es). 



\subsection{Motivations of NaaS}
\label{sec:whyNaaS}

The main motivation of NaaS is the desire to address the inefficiencies of traditional cloud networking.  In recent times, it is widely believed that (traditional) networking is a bottleneck in cloud innovations due to its reliance on \emph{manual configurations} as a result of \emph{tight coupling of services and infrastructure}. Traditional DC networking solutions do not scale since they require manual configuration (which can increase commissioning time to days and weeks). In line with cloud computing convention, it is desirable to have instant self-service provisioning. NaaS is also motivated by the desire to avoid the expensive process of application rewriting which can result with limitations of traditional networking (such as the lack of broadcast domain abstraction, or the lack of support for custom assigned IPs to the virtual servers). The advancements of virtualization technology, the rise of high-level APIs for automated provisioning and service orchestration, and the economy of operating at cloud scale makes NaaS a very attractive proposition for both the CSPs and the service consumsers. NaaS is also motivated by the desire to utilize the available compute capacity in DCs to implement networking functions typically implemented on  \emph{middleboxes}\footnote{Middleboxes are intermediate devices (such as load balancers, firewalls, WAN optimization, intrusion detection systems, etc.) typically implemented on proprietary physical hardware.} as virtualized cloud instances to offset the expensive proposition of conventional proprietary middleboxes.

\subsection{Benefits of NaaS}
\label{sec:benefitsNaaS}
 
The prime benefit of NaaS is the \emph{i)} \emph{agility in deployment} with which networks can be provisioned in a matter of minutes and the \emph{ii)} \emph{scalability} and \emph{elasticity} of service using which the network can be upgraded as and when desired with the convenient pricing model of pay-as-you-use. The users of NaaS have the benefit of \emph{iii)} full \emph{automation} using which the consumer can program, manage, and orchestrate the network with programmatic control at convenient granular utility pricing leading to \emph{iv)} \emph{savings} and \emph{productivity}. The consumer can support \emph{v)} \emph{custom policies} in its own virtual network which leads to \emph{vi)} enterprise network \emph{innovation}. This can mean orchestration of several networking functions as desired by the user such as custom routing, load balancing, network isolation, firewalling, custom addressing, etc. In addition, the consumers can \emph{vii)} work with any hardware, or even a mixed vendor hardware, without worrying about the burden of configuration and management and thereby \emph{avoid vendor lock-in}. The CSPs also gain a lot from offering NaaS: e.g., by utilizing the features of \emph{viii)} \emph{VM mobility}, the DC resources can be effectively utilized by moving the VMs from loaded servers to idle servers without disrupting the VNs. The NaaS model has full support for \emph{ix)} \emph{multi-tenancy with isolation} which leads to better economics for CSPs while satisfying the isolation and security requirements of customers. Lastly, another benefit of the NaaS approach is \emph{x)} \emph{fault-localization} with which a fault in one VN (due to policy configuration or otherwise) does not cascade to affect the whole network infrastructure.

\vspace{2mm}
\section{Traditional Network Virtualization Techniques}
\label{sec:traditional}
\vspace{2mm}

In this section, we will discuss various network virtualization techniques. Network virtualization is not a new concept having existed in many different independent guises \cite{chowdhury2010survey}. In particular, numerous solutions have been proposed to address aspects of the network virtualization \cite{koponen2014network} including: \emph{i)} \emph{VLAN technology} for virtualizing Ethernet LANs by introducing the subnet abstraction; \emph{ii)} \emph{NAT technology} for virtualization the IP address space by allowing multiple enterprises to share the same private IP space; \emph{iii)} \emph{virtual-circuit technology} (such as MPLS) which virtualizes the path by allowing multiple virtual circuits to share the same physical circuit; \emph{iv)} \emph{VPN technology} for virtualizing a public network such that multiple virtual private networks can share the infrastructure; \emph{v)} \emph{VRF technology} for virtualizing the router by allowing virtual forwarding tables in your router or even run routers within VMs; \emph{vi)} \emph{overlay networks} for building a virtual logical network on top of another physical network using tunneling mechanisms.  We will next discuss two traditional techniques, VLANs and overlays, which are arguably more important than others in the context of NaaS. 



\subsection{Virtual Local Area Networks (VLANs)}
\label{sec:VLAN}


VLANs define a mechanism for partitioning a physical L2 network into several virtual LANs each with its own broadcast domain and traffic isolation. With VLANs, switching is based on destination MAC as well as the VLAN ID (tag value) which results in simpler design of L2 networks as the physical location of a node does not dictate its membership. 
IEEE 802.1Q employs the use of VLAN tagging to specify VLAN memberships. With the VLAN ID field defined to be 12 bit, there is a strict limit of 4096 VLANs on a single Ethernet network which worked fine in the pre-cloud era but is grossly insufficient now as modern clouds greatly rely on virtualization. Some other important deficiencies of VLAN technology in the context of cloud computing and multi-tenant DCs include the lack of address space isolation (which is problematic for multi-tenant DCs) and \emph{`equal-cost-multipath'} (ECMP) support (which is inefficient from the DC operator's point of view) \cite{Jain2013VN}.

\subsection{Overlay Networks using Tunnels}
\label{sec:overlay}
 
An overlay network is essentially a virtualized logical network built on top of a physical network with tunnels interconnecting edge devices. The overlay network is typically decoupled from the underlying physical network through dual address spaces representing a tunnel encapsulation with the virtual address space on the inside and the physical address space on the outside. The overlay network appears to the nodes connecting to it as a native network with the possibility of multiple overlays existing on the same underlying physical infrastructure (which allows support for multi-tenancy). Each overlay network is effectively a distinct logical network which can support service properties such as an arbitrary policy of L2, L3, access control list (ACL) processing distinct from the physical network. This makes overlay networking a popular technique for supporting disruptive innovations in networks without requiring interventions in the core network \cite{anderson2005overcoming}. Overlay techniques are popular in building NaaS solutions using the \emph{`virtual overlay network'} (VON) model which utilizes virtual switches that reside on the edges of a DC network. 


\section{Enabling Technologies of NaaS}
\label{sec:EnablingTechnologies}

\vspace{2mm}
\subsection{Everything-as-a-service (XaaS)} 

Cloud services has traditionally been defined in a three tiered hierarchy differentiated by the level of abstraction presented to the service user \cite{armbrust2010view}. The lowest tier is \emph{`infrastructure-as-a-service'} (IaaS) in which the CSP provides a complete computing infrastructure to consumers in the form of VMs and servers that the end users can modify according to their needs. The next tier is \emph{`platform-as-a-service'} (PaaS) in which the CSP provides a integrated development platform capable of supporting the complete life-cycle of building and delivering applications and services over the web. The upper tier, which is the tier most visible to end-users and the most abstract, is \emph{`software-as-a-service'} (SaaS) which runs on top of the PaaS layer. SaaS users can access software services via the cloud without bothering about hardware and software implementation details. There is now a significant interest in having cloud services diversify beyond the traditional three-tiered services model and embrace an all-encompassing \emph{`everything-as-a-service'} (XaaS) model in which every conceivable IT facility---ranging from computation, storage, data, platform, infrastructure, software as well as \emph{networking functions}---will be offered as a service utility computing style. The benefits of XaaS includes broadening the domain of services that can reap the benefits of cloud computing (such as a lower barrier to entry in deployment, reduced CAPEX and OPEX, massive scalability, support for multi-tenancy as well as independence from constraints of being bound to device and location). For a more detailed exposition of XaaS and cloud computing, please refer to references \cite{armbrust2010view} \cite{qadir2014prog}.

\vspace{2mm}
\subsection{Software Defined Networking (SDN)}  
  
Although, the traditional internetworking architecture, developed in 1970s, forming the basis of the Internet has remarkably survived for more than 40 years, evolving the Internet ecosystem has not been easy. The lack of suitable control abstractions in the original Internet design choice---in particular, the lack of well-defined standard based interface between the data and control planes---has made the Internet architecture resistant to change. Without suitable abstractions in place, vertically integrated systems have became the architectural norm leading to problems of vendor lock-in and impeded innovation. The lack of suitable abstractions for programming the network as a whole has meant that supporting cloud-era applications is difficult with undesirable burden of manually configuring various network switches through vendor-specific command-line-interfaces (CLIs)---a process that is cumbersome and error prone.

Following in the footsteps of software-defined radio (SDR) technology, which enabled programmability of wireless radios, there is now enormous interest in creating programmable software-defined networks \cite{qadir2014prog}. SDN technology, along with cloud computing, is helping in creating an exciting new vista of opportunities for networking innovations \cite{qadir2014acmfit}. 

\begin{figure}[!ht]
\centering
\subfigure[In traditional networking, the control planes (CP) and the data planes (DP) are co-located on devices to ensure decentralized network control.]{
 \includegraphics[width=.35\textwidth]{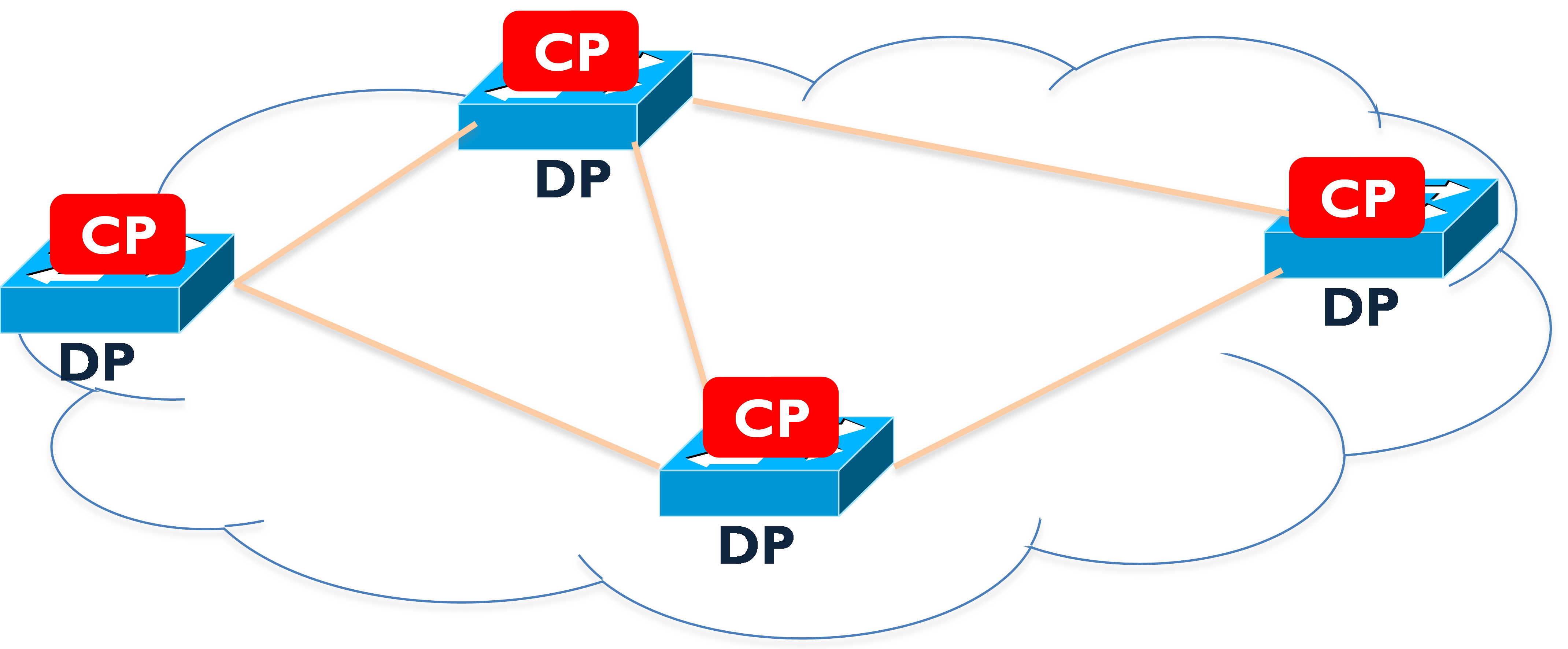}
   \label{fig:PreSDN}
 }
\subfigure[In SDNs, the DPs and CPs are separated with a centralized controller controlling multiple DPs while supporting a \emph{southbound} API to the DPs and a \emph{northbound} API to the SDN applications.]{
   \includegraphics[width=.45\textwidth]{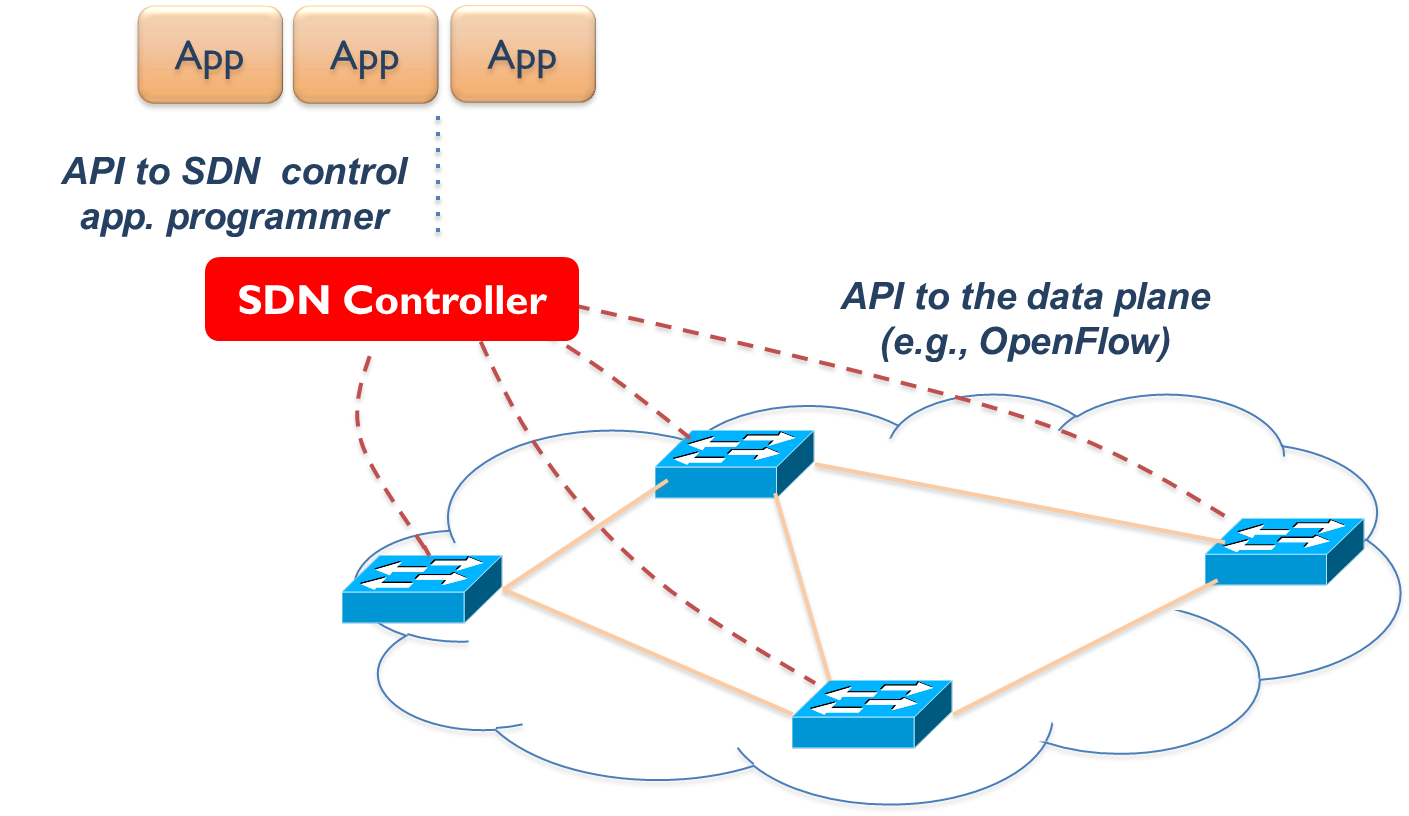}
   \label{fig:SDN}
}
\caption{Comparison of Traditional and SDN network architectures}
\label{fig:SDNoverall}
\end{figure}

The main insight of SDNs is to allow horizontally integrated systems by allowing the separation of the control plane and the data plane \cite{mendoncca2013survey} (see figure \ref{fig:SDNoverall} for an illustration) while providing increasingly sophisticated set of abstractions. SDN has revolutionized the networking industry by providing architectural support for ``programming the network''. SDN promises to be a major paradigm shift in networking landscape leading to improved and simplified networking management and operations. 


The development of SDNs is supported by a burgeoning open-source community. The Open Network Foundation (ONF) oversees the standardization efforts of SDNs. Broadly speaking, there are two main classes of APIs in the SDN architecture: \emph{i)} the Southbound API defines an interface between a centralized network controller\footnote{The centralized SDN network controller can itself be built as a distributed system to be scalable and avoid a single point of failure.} and networking devices \cite{mckeown2008openflow}, while \emph{ii)} the Northbound API defines the interface exposed by the controller to the network applications. \emph{OpenFlow} \cite{mckeown2008openflow} is an example standard southbound API which has been standardized by the ONF. With the control logic implemented in a separate controller, and a standardized control API between the controller and the data planes, the vision of programming the network using a high-level control language can be achieved. With the separation of the control plane from the data plane, it is possible for third party/ open-source developers to write program applications for the controller. This allows networks to employ programmable commodity hardware rather than `closed' vendor hardware, increasing flexibility and development while reducing costs.

An initial SDN use case, espoused in \cite{mckeown2008openflow}, was allowing researchers to run experimental protocols in virtualized `\emph{slices}' of the production network. The concept of slicing network through virtualization technology predates SDN, and has been used in the PlanetLab and the Emulab projects, and more recently in the NSF funded global environment for network innovations (GENI) project. Taking this further, the concept of a `\emph{network hypervisor}' has recently been proposed to virtualize the network's forwarding plane. 
The network hypervisor implements a network-wide software layer through which it supports multiple virtualized networks which are decoupled from their underlying hardware instantiation. The concept of network hypervisor is analogous to the conventional hypervisor concept that refers to a VM monitor that runs as a host program on a physical machine and controls the potential multitude of VMs on that machine. 


\subsection{Network Virtualization}

Network virtualization technology is utilized in NaaS settings for \emph{i)} for virtualizing network functions in the cloud, and \textit{ii)} for virtualizing networks to provide the VN abstraction. We discuss these two respective directions next. 


\vspace{2mm}
\subsubsection{Network Functions Virtualization (NFV)}
\label{sec:nfv}

\begin{table*}[!ht]
\tiny
\centering
\caption{Summary of the main enabling technologies of NaaS}
\label{tab:enablingTech}
\begin{tabular}{|
>{\columncolor[HTML]{EFEFEF}}m{1.5cm} |m{2cm}|m{3cm}|m{3cm}|m{1.5cm}|}
\hline
\multicolumn{1}{|l|}{\cellcolor[HTML]{FFFFFF}{\color[HTML]{333333} }}                                    & \cellcolor[HTML]{EFEFEF}\textbf{XaaS}                                                                            & \cellcolor[HTML]{EFEFEF}\textbf{SDN}                                                                                                                                                              & \multicolumn{1}{l|}{\cellcolor[HTML]{EFEFEF}\textbf{NFV}}                                                                                                      & \multicolumn{1}{l|}{\cellcolor[HTML]{EFEFEF}\textbf{VON}}                                      \\ \hline
{\color[HTML]{333333} \textit{Reason of existence}}                                                    & To reap the efficiency of centralized utility computing                                                          & To `open' up the ossified networking landscape that had vertically integrated architectures by emphasizing the separation of control and data planes and their communication using open interfaces & To cloudify networking functions by running services typically running in  dedicated proprietary hardware as virtualized software-based functions in the cloud & To accelerate service provisioning and orchestration especially in the DC environments \\ \hline
{\color[HTML]{333333} \textit{\textbf{Initial Market}}}                                                  & Small enterprises and businesses                                                                                 & Campus, DC, and cloud environments                                                                                                                                                        & Telecom service provider environment                                                                                                                           & Campus, DC, and cloud environments                                                     \\ \hline
{\color[HTML]{333333} \textit{\textbf{Target Devices}}}                                                  & \multicolumn{4}{c|}{Commodity servers and switches (supporting open non-proprietary software and protocols)}                                                                                                                                                                                                                                                                                                                                                                                                                                                                           \\ \hline
{\color[HTML]{333333} \textit{\textbf{\begin{tabular}[c]{@{}c@{}}Initial \\ Applications\end{tabular}}}} & IT outsourcing, Infrastructure, platform, and software as a service                                              & Enterprise security, service orchestration and provisioning in clouds                                                                                                                             & Software-based virtualized implementations of telecom functions such as routers, load balancers, firewalls, functions of cellular networks, etc.               & Service orchestration and provisioning in clouds                                               \\ \hline
{\color[HTML]{333333} \textit{\textbf{Protocols Used}}}                                                  & SDN and VON                                                                                                      & Openflow, PCEP, BGP, NETCONF, SNMP, etc.                                                                                                                                                          & None yet                                                                                                                                                       & Tunneling protocols such as VXLAN, NVGRE, and STT                                              \\ \hline
{\color[HTML]{333333} \textit{\textbf{Solutions}}}                                                       & Commercial offerings from companies such as Amazon, Google, etc.; Open source software OpenStack also available. & Numerous projects including open source controllers NOX, POX, Floodlight, and open source virtual switches such as Open vSwitch                                                                    & New projects are emerging such as CloudNAV                                                                                                                     & VMWare's NSX, PlumGrid, Midokuro, Nuage, etc.                                                  \\ \hline
{\color[HTML]{333333} \textit{\textbf{\begin{tabular}[c]{@{}c@{}}Formal \\ Leadership\end{tabular}}}}    & None                                                                                                             & Open Networking Foundation                                                                                                                                                                        & ETSI NFV working group                                                                                                                                         & None                                                                                           \\ \hline
\end{tabular}
\end{table*}

\begin{figure}[!ht]
\begin{center}
\includegraphics[width=.47\textwidth]{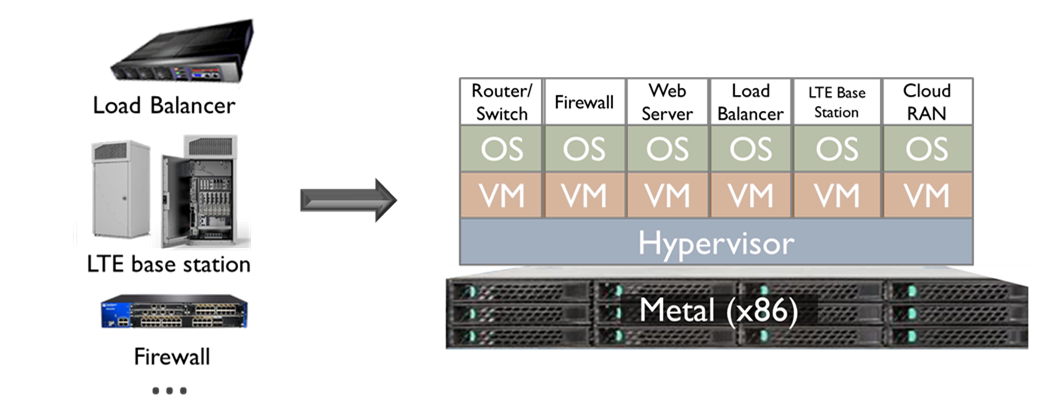}
\caption{Network functions virtualization (NFV) is used to convert fixed function hardware network appliances into virtualized cloud software instances that run on commodity infrastructure hosted in cloud DCs.}
\label{fig:nfv}
\end{center}
\end{figure}

Motivated by the success of SDN-based network virtualization and cloud computing, there is great interest in the telecom service provider community to decouple the functionality of telecom devices and services (such as mobile network node, radio network controller, etc.) from dedicated hardware and enable ``\emph{network functions virtualization}'' (NFV) by converting fixed function hardware network appliances into virtualized cloud software instances (as illustrated in figure \ref{fig:nfv}). The genesis of the NFV efforts is recent with leading network service providers (such as AT\&T, British Telecom, Deutsche Telecom, etc.) forming an industry specification group under the aegis of European Telecommunications Standards Institute (ETSI) in 2012. The NFV group aims to leverage standard cloud and virtualization technology for decoupling network functions from proprietary hardware devices. 

NFV technology also allows the providers to make the data plane programmable which will facilitate in orchestrating middlebox functionality efficiently. Traditionally, it is not uncommon for a single packet to undergo processing in the data plane through multiple middleboxes that are used to augment data plane processing by L2 and L3 switches. This functionality of network infrastructure orchestration can be implemented in the NFV framework by \emph{`service chaining'} through virtual network functions (such as ACLs, load balancing, etc.) running as virtualized cloud instances in DCs. The concept of NFV extends to any data plane packet processing and control plane function in mobile or fixed networks including, but not limited to, mobile network nodes and traditional switching devices such as routers, switches, home gateways, etc.

NFV brings many of the same advantages that SDN offers---\emph{i) virtualization:} using resources regardless of where it is physically located; \emph{ii) orchestration:} managing thousands of devices through an API; \emph{iii) programmability:} the ability to change behavior dynamically on the fly---to the telecom world. In a few years time, carrier and telecom networks will increasingly emulate DCs and clouds in their reliance on commodity hardware, virtualization technology, and open software and interfaces in a break from the current scenario in which telecom service providers are full of proprietary vertically integrated hardware appliances. Apart from the NFV vision of `cloudifying' middlebox services and network functions, keeping in mind that middleboxes are typically deployed at the network's edge, there is also great interest in assimilating the middlebox functionality into the SDN framework at the edge in software. The two trends of SDN and NFV, although independent, can coexist and complement each other in implementing the vision of NaaS. 


\vspace{2mm}
\subsubsection{Virtualized Overlay Networks (VONs)}
\label{sec:VON}

The `holy grail' of the cloud computing paradigm is the vision of installing a generic `network fabric' which can be then automatically programmed to provide any service without any need of manual configuration of core network nodes. An emerging architecture that promises to fulfill this vision is to have a protocol agnostic network fabric or network core which is focused only on IP transport along with a hypervisor overlay network---known as a VON---which interconnects virtualized software switches running at the edge (on commodity x86 hardware rather than on ASICs) in which the advanced network functionality is implemented totally in software. In the context of NaaS, VONs play a large role in facilitating the creation of the VN abstraction allowing much of the networking functionality to be recreated in software at the edges in a totally virtualized fashion.

 \begin{figure}[t]
\begin{center}\emph{•}
\includegraphics[width=.5\textwidth]{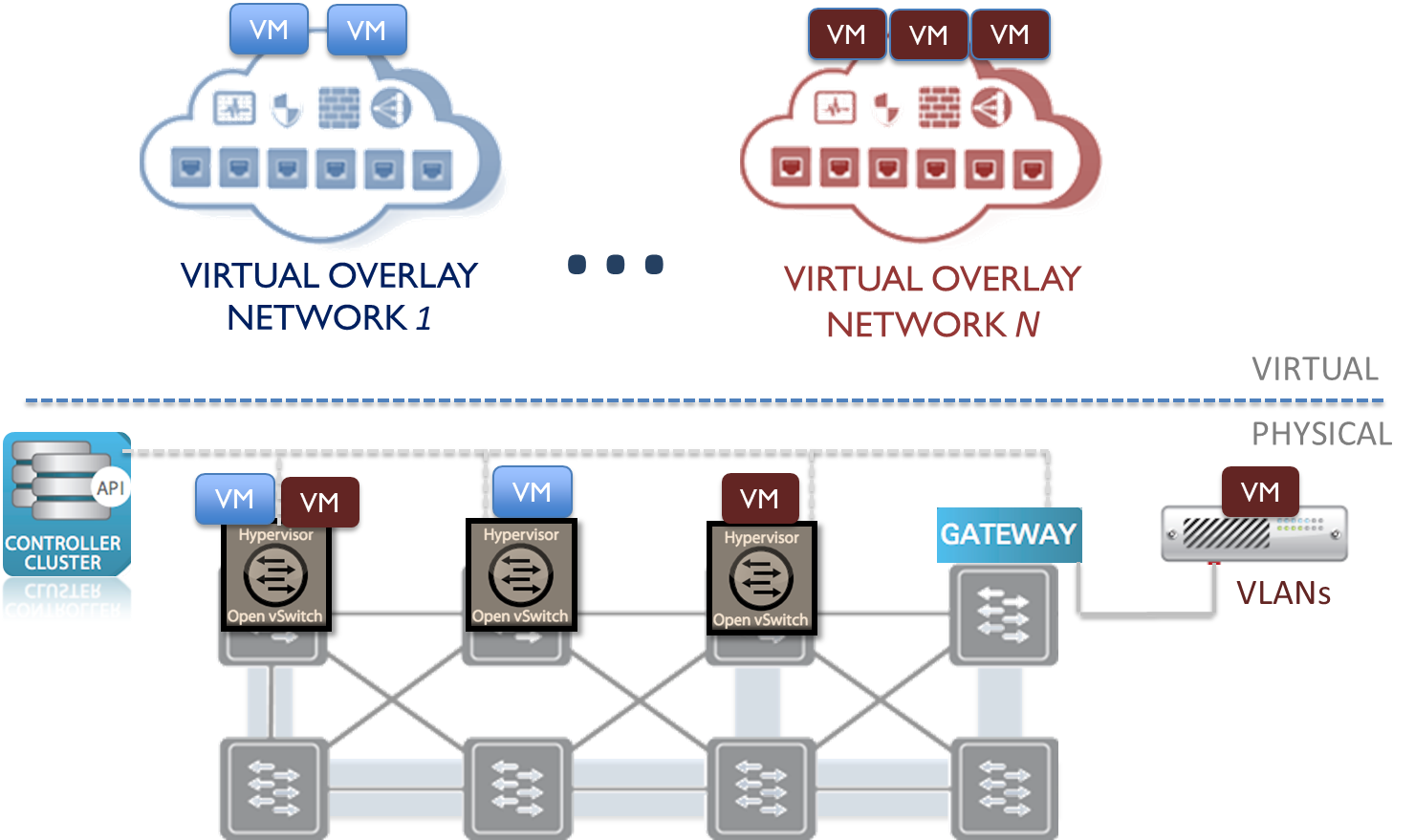}
\caption{The concept of Virtualized Overlay Network (VON).}
\label{fig:von}
\end{center}
\end{figure}

The trend of `\emph{virtualized} overlay networking' is driven by increasing virtualization of networking. With the estimated number of physical ports being overtaken by virtualized ports, we are currently observing a significant inflection point in networking history with major architectural implications. In particular, an hypervisor overlay with a networking fabric constructed out of SDN technology (the SDN/ VON hybrid architecture) can become the modern functional equivalent of the traditionally influential end-to-end principle. In this new SDN/ VON architecture (illustrated in figure \ref{fig:von}), the SDN based fabric will be the equivalent of the traditional network core (focused on hardware based rapid switching), while the hypervisor switches, such as the \emph{Open vSwitch} \cite{pfaff2009extending}, will constitute the new (software-based) edge devices. In the future, it is anticipated that the SDN/ VON hybrid architecture will also subsume the functionality of MPLS and middleboxes to offer a clean split between the core and the edge.  Already, software switches are supported in hypervisors to allow VMs to communicate. The great promise of VONs is to allow vSwitches to be connected over a protocol agnostic network fabric with the vSwitches able to support arbitrary protocols over the VON. This paradigm shift to software control fundamentally changes the pace of innovation, and opens up a world of new possibilities. 

The architecture of a VON is illustrated in figure \ref{fig:von} with a fabric comprising hardware switches and end hosts with software vSwitches. Since it is possible for VLANs to span several DCs connected via L3 networks, simulating a L2 network requires the construction of an overlay constructed using tunneling protocols. It is typical for modern VON approaches to utilize SDN to manage the control plane to facilitate cloud-scale rapid provisioning of the VN abstraction. In this fashion, the network provisioning, intelligence is at the edge in virtualized switches and in software and does not require any change in the underlying infrastructure. 


Tunneling (or encapsulation) protocols are an important component of VON based NaaS solutions. A number of tunneling protocols including `virtual extensible LAN' (VXLAN), `network virtualization using generic routing encapsulation' (NVGRE), `stateless transport tunneling' (STT), and `generic network virtualization encapsulation' (Geneve)---all IETF draft standards---have been proposed to the IETF's Network Virtualization over L3 (NVO3) working group. These protocols address the inefficiencies associated with traditional solutions such as VLANs in multi-tenant DCs by enabling the VN abstraction using which---assuming appropriate control plane support such as those provided by SDN---VNs can be created cloned, copied, and moved just like VMs. 


In summary, VONs can be used to support existing IP transport infrastructures without modification, and for supporting new protocols (such as OpenFlow) without requiring hardware changes in the network fabric. These benefits make VON a very popular NaaS enabling solution. It is worth highlighting that most of the current NaaS solutions from vendors such as Nicira (VMware), PlumGrid, IBM, BigSwitch, etc. adopt VONs in their design.

\section{Conclusion}
\label{sec:conclusion} 

NaaS is a framework that can be implemented on existing networking infrastructures of cloud to not only overcome its limitations but also enhance its capabilities to a great extent. By offering NaaS, cloud service providers can simultaneously improve their infrastructural utilization while introducing agility in deployment, provisioning and orchestration. In this article, we have discussed some key enabling technologies of NaaS: XaaS, SDN, NFV, and VONs. XaaS is about reaping the efficiency and flexibility of cloud-computing's utility-computing paradigm. SDN makes the control plane programmable by emphasizing a separation of the control and data planes and introducing new control abstractions. NFV is about making the data plane programmable by allowing the implementation of networking middleboxes to run as virtualized cloud functions. VON decouples the network functionality from its realization in the physical network at the edge of the network in software (without requiring any change in the core network). These technologies share the objectives of introducing openness, innovation, and efficiency and promise the ability to efficiently offer NaaS services. While the particular niche of these technologies are distinct, and they can be plausibly implemented independently of each other, they interwork very well and synergize to strengthen their mutual value proposition. These technologies provide today the technological underpinnings of a disruptive NaaS service that promises to revolutionize networking.

\section{Acknowledgement}
\label{sec:conclusion} 

The research work presented in this paper is conducted as part of the Mobile Cloud Networking project, funded from the European Union Seventh Framework Program under grant agreement n°[318109].


\begin{IEEEbiographynophoto}{Junaid Qadir}
Junaid Qadir is an Assistant Professor at the School of Electrical Engineering and Computer Science (SEECS) at the National University of Sciences and Technology (NUST). He is the director of the Cognet lab at SEECS, NUST. His research interests include wireless networks, programmable networks, and cognitive radio networks. He is a senior member of IEEE.
\end{IEEEbiographynophoto} 

\begin{IEEEbiographynophoto}{Nadeem Ahmed}
Nadeem Ahmed is an Assistant Professor at the School of Electrical Engineering and Computer Science (SEECS) at the National University of Sciences and Technology (NUST). His research interests include wireless sensor networks, software-defined networks, and cloud computing.
\end{IEEEbiographynophoto} 

\begin{IEEEbiographynophoto}{Faqir Zarrar Yousaf}
Faqir Zarrar Yousaf he is working as a Research Scientist at NEC Laboratories Europe in Heidelberg, Germany where the core focus of his research is in the area of User Plane congestion management, application level scheduling and network function virtualization. He completed his PhD from Dortmund University of Technology (TU Dortmund), Germany in April 2010.
\end{IEEEbiographynophoto} 

\begin{IEEEbiographynophoto}{Ali Taqweem}
Ali Taqweem completed his Bachelors in Electrical Engineering from the School of Electrical Engineering and Computer Science (SEECS) at the National University of Sciences and Technology (NUST). He was an internee in the Cognet lab in 2013.
\end{IEEEbiographynophoto} 
    
\bibliography{ref}
\end{document}